\documentclass[aps,prd,preprintnumbers,groupedaddress,nofootinbib,amssymb,notitlepage,eqsecnum]{revtex4-2}
\usepackage{here}
\usepackage[dvipdfmx]{graphicx}
\usepackage{amsmath,amsthm,amssymb}
\usepackage{bm}
\usepackage{color}

\usepackage{amsfonts}
\usepackage{dcolumn}
\usepackage{hyperref}
\allowdisplaybreaks[1]
\usepackage{stackengine}

\newcommand{\be}{\begin{equation}}  
\newcommand{\ee}{\end{equation}}
\newcommand{\ba}{\begin{eqnarray}}
\newcommand{\ea}{\end{eqnarray}}

\newcommand{\rd}{{\rm d}}

\newcommand{\bem}{\begin{bmatrix}}
\newcommand{\eem}{\end{bmatrix}}
\newcommand{\Mpl}{M_{\rm Pl}}

\allowdisplaybreaks

\begin{document}

\preprint{YITP-23-90, WUCG-23-08}

\title{Stability of Schwarzshild black holes in quadratic gravity \\
with Weyl curvature domination}

\author{Antonio De Felice$^{1}$ and Shinji Tsujikawa$^{2}$}

\affiliation{
$^1$Center for Gravitational Physics and Quantum Information, 
Yukawa Institute for Theoretical Physics, Kyoto University, 
606-8502, Kyoto, Japan\\
$^2$Department of Physics, Waseda University, 3-4-1 Okubo, 
Shinjuku, Tokyo 169-8555, Japan}

\begin{abstract}

We study the linear stability of static and spherically symmetric (SSS)
black holes (BHs) in the presence of a Weyl-squared curvature 
besides an Einstein-Hilbert term in the action. 
In this theory, there is always an exact Schwarzschild BH 
irrespective of the Weyl coupling constant $\alpha$, 
with the appearance of a non-Schwarzschild solution 
for a particular range of the coupling of 
order $|\alpha| \approx r_h^2$ (where $r_h$ is 
the horizon radius).
On the SSS background, we show that the propagating degrees 
of freedom (DOFs) are three in the odd-parity sector and four 
in the even-parity sector. Since the number of total seven DOFs coincides with those on the Minkowski and isotropic cosmological backgrounds, the Weyl gravity does not pose a strong coupling problem associated with the vanishing kinetic term of dynamical perturbations.
The odd-parity perturbations possess at least one ghost mode, 
but the propagation speeds of all three dynamical modes 
are luminal.
In the even-parity sector, our analysis, based on the WKB approximation, shows that, besides the appearance of at least one ghost mode, the Schwarzschild solution is prone to both radial and angular Laplacian instabilities of several dynamical perturbations for the Weyl coupling in the range $|\alpha| \gg r_h^2$.
For large radial and angular momentum modes, the time scales of such instabilities are much shorter than the horizon distance $r_h$ divided by the speed of light.
In the coupling regime $|\alpha|\lesssim r_h^2$, the WKB approximation does not hold any longer, and a different analysis should be performed 
if one wants to state the stability of both the Schwarzschild and 
non-Schwarzschild BH solutions in this range of model parameters.

\end{abstract}

\date{\today}

%\pacs{04.50.Kd, 95.36.+x, 98.80.-k}

\maketitle

%%%%%%%%%%%%%%%%%%%%%%%%%%%%%%%%%%%%%%%%%%
\section{Introduction}
\label{introsec}
%%%%%%%%%%%%%%%%%%%%%%%%%%%%%%%%%%%%%%%%%%

Black holes (BHs) are fundamental objects arising as vacuum 
solutions in General Relativity (GR). 
A static and spherically symmetric (SSS) vacuum configuration in GR gives rise to a Schwarzschild solution characterized by a horizon radius $r_h$.
In theories beyond GR, it is possible to realize non-Schwarzschild BH solutions whose geometries are modified by the presence of additional 
degrees of freedom (DOFs). 
In scalar-tensor or vector-tensor theories, for example, there are some asymptotically-flat BH solutions endowed with scalar or vector hairs \cite{Kanti:1995vq, Torii:1996yi, Chen:2006ge, Guo:2008hf, Pani:2009wy, Kleihaus:2011tg, Sotiriou:2013qea, Sotiriou:2014pfa, Ayzenberg:2014aka, Maselli:2015tta, Kleihaus:2015aje, Doneva:2017bvd, Silva:2017uqg, Antoniou:2017acq, Heisenberg:2017xda, Heisenberg:2017hwb, Minamitsuji:2018xde, Silva:2018qhn, Langlois:2022eta, Minamitsuji:2022vbi, Minamitsuji:2022mlv, DeFelice:2022qaz, Aoki:2023jvt, Tsujikawa:2023egy}. 
With the advent of gravitational astronomy \cite{LIGOScientific:2016aoc}, 
we can now probe the physics of strong gravity regimes and the possible deviations from GR \cite{Berti:2015itd, Barack:2018yly, Berti:2018cxi}.

GR is described by an Einstein-Hilbert action with the Lagrangian $\Mpl^2 R/2$, where $\Mpl$ is the reduced Planck mass and $R$ is the Ricci scalar. On strong gravitational backgrounds, it is expected that 
quadratic curvature terms may modify the spacetime structure and dynamics. The general quadratic curvature contributions to the action consist of the terms $R^2$, $R_{\mu \nu}R^{\mu \nu}$, and $R_{\mu \nu \rho \sigma}R^{\mu \nu \rho \sigma}$, where $R_{\mu \nu}$ is the Ricci tensor and $R_{\mu \nu \rho \sigma}$ is the Riemann tensor. Since the Gauss-Bonnet curvature invariant ${\cal G}=R^2-4R_{\mu \nu}R^{\mu \nu}
+R_{\mu \nu \rho \sigma}R^{\mu \nu \rho \sigma}$ is a topological term that does not affect the spacetime dynamics in four dimensions, the general gravitational action up to quadratic-order curvature terms 
is given by 
\be
{\cal S}=\frac{\Mpl^2}{2} \int {\rm d}^4 x 
\sqrt{-g} \left( R-\alpha C_{\mu \nu \rho \sigma} 
C^{\mu \nu \rho \sigma}+\beta R^2 \right)\,,
\label{actiontotal}
\ee
where $g$ is a determinant of the metric tensor $g_{\mu \nu}$, $\alpha$ and $\beta$ are coupling constants, and $C_{\mu \nu \rho \sigma}$ is the Weyl tensor whose squared is given by 
\be
C_{\mu \nu \rho \sigma} C^{\mu \nu \rho \sigma}
=2R_{\mu \nu} R^{\mu \nu}-\frac{2}{3}R^2+{\cal G}\,.
\label{Weyl}
\ee
If we regard the higher-order curvature terms as one-loop quantum corrections, the theory is renormalizable at the price of having ghost DOFs induced by the Weyl term \cite{Stelle:1976gc}. 

In theories without the Weyl curvature, i.e., $\alpha=0$ in Eq.~(\ref{actiontotal}), there exists one healthy scalar mode (``scalaron'') for $\beta>0$ besides two tensor polarizations. On the cosmological background, the scalaron potential induced by the Lagrangian $\beta R^2$ can derive an accelerated expansion of the Universe \cite{Starobinsky:1980te}. If we apply the same theory to 
BH physics, it is known that there are no hairy SSS solutions 
other than the Schwarzschild solution. This is related to the fact that $f(R)$ gravity is equivalent to Brans-Dicke theories \cite{Brans:1961sx} with a scalar potential arising from a nonlinear function of $f(R)$ \cite{OHanlon:1972xqa, Chiba:2003ir, DeFelice:2010aj}. 
In such theories, several authors showed that the no-hair property holds for the SSS BHs \cite{Hawking:1972qk, Bekenstein:1995un, Sotiriou:2011dz, Faraoni:2017ock, Minamitsuji:2022mlv}.

In the presence of the Weyl term, the Schwarzschild BH is an exact solution on the SSS background for any arbitrary couplings $\alpha$. 
When the coupling $|\alpha|$ is of order $r_h^2$, it is known that the other asymptotically-flat non-Schwarzschild branch appears 
besides the Schwarzschild branch \cite{Lu:2015cqa, Lu:2015psa}.
For $|\alpha|$ exceeding the order of $r_h^2$, there is only the Schwarzschild branch. In this coupling regime, the Weyl squared dominates over the Einstein-Hilbert term around the horizon. 
For $\alpha>0$, the intersection of Schwarzschild and non-Schwarzschild branches occurs at the point $r_h/\sqrt{2\alpha} \simeq 0.876$ \cite{Lu:2015cqa, Lu:2015psa, Kokkotas:2017zwt, Podolsky:2019gro, Huang:2022urr}. The non-Schwarzschild solution is present up to the value $r_h/\sqrt{2\alpha} \simeq 1.143$, above which the BH mass becomes negative. The non-Schwarzschild branch exists in the range $0.876 \le r_h/\sqrt{2\alpha} \le 1.143$, while the Schwarzschild branch is 
present for any positive values of $r_h/\sqrt{2\alpha}$.

In theories with $\beta=0$ in Eq.~(\ref{actiontotal}), the analysis of linear perturbations on the Minkowski and isotropic cosmological backgrounds shows that there are seven dynamical DOFs (four tensor, two vector, one scalar modes) in total \cite{Hindawi:1995an, Bogdanos:2009tn, Deruelle:2010kf, Hinterbichler:2015soa, Salvio:2018crh}.
Apart from the two massless tensor modes, the Weyl term generates a mass squared $m_W^2=1/(2\alpha)$ for the other five DOFs on the Minkowski background. If $\alpha>0$, the tachyonic instabilities of massive 
modes do not arise, but there are five ghosty propagating DOFs. 
Meanwhile, the Laplacian instabilities of such massive modes are absent, so they can be regarded as ``soft ghosts'' \cite{Smilga:2004cy}
at the classical level around the Minkowski vacuum. This should not be the case if the ghosts are coupled to other fields \cite{Zeldovich:1977vgo, Vilenkin:1985md, Cline:2003gs, Gorbunov:2010bn}. 

The presence of Weyl ghosts on the Minkowski background implies that these new modes may lead to some instabilities on other curved backgrounds. 
In this paper, we would like to address the propagation and linear 
stability of dynamical perturbations on the SSS background. To extract the effect of the Weyl curvature term on the dynamics of perturbations, we study theories given by the action (\ref{actiontotal}) with $\beta=0$. If the number of propagating DOFs is less than seven on a particular background, this implies the presence of a strong coupling problem. This problem arises in some other higher-curvature gravity theories such as Einsteinian cubic gravity \cite{Bueno:2016lrh}, where the propagating DOFs on the maximally symmetric background are smaller than those around general curved backgrounds \cite{Pookkillath:2020iqq, BeltranJimenez:2020lee, DeFelice:2023vmj, Jimenez:2023esa}. In Weyl gravity, we will show that there are seven dynamical DOFs (three in the odd-parity sector and four in the even-parity sector) for perturbations on the SSS background. Although there are ghosts in both odd- and even-parity sectors, the theory does not give rise to the strong coupling problem.

To study the linear stability of BHs, we will exploit the WKB approximation in which the solution to perturbations is dominated by the large angular frequency and high radial and angular momenta. The WKB prescription loses its validity in the massive regime 
where the Weyl mass squared $m_W^2=1/(2\alpha)$ provides 
non-negligible contributions to the solutions of perturbations outside the horizon. To avoid the breakdown of the WKB approximation in the vicinity of the horizon, we require the condition 
$|m_W^2| r_h^2 \ll 1$, i.e., $|\alpha| \gg r_h^2$. This is the regime in which the hairy non-Schwarzschild branch is absent. However there exists the Schwarzschild branch, so we will use this background solution for studying the BH stability. This does not generally mean that the 
hairy non-Schwarzschild branch is stable. It just implies that we cannot apply the WKB approximation to accommodate their stability and 
that another study is necessary to clarify this issue.

We note that, for the monopole mode (multipole $l=0$) 
with a negligible radial 
wavenumber $k$ relative to mass terms, a long-wavelength instability analogous to the Gregory-Laflamme instability \cite{Gregory:1993vy} was reported in Ref.~\cite{Held:2022abx} for the Schwarzschild BH in the coupling range $|\alpha| \gg r_h^2$.
For $l=0$, the number of dynamical DOFs can be different and reduced in comparison to the modes $l \geq 2$ (which is the case 
for other modified gravity theories \cite{Kobayashi:2014wsa, Kase:2020qvz, Kase:2021mix, Kase:2023kvq}).
In this paper, we wish to clarify whether there are Laplacian instabilities of the Schwarzschild BH for large radial and angular momentum modes ($k r_h \gg 1$ and $l \gg 1$), by properly dealing with all dynamical perturbations under the WKB approximation.

We will show that, albeit with the appearance of at least one ghost mode, all of the odd-parity dynamical perturbations have luminal propagation speeds in both radial and angular directions. Hence the Laplacian instabilities in the odd-parity sector are absent at the classical level. 
In the even-parity sector, besides the appearance of at least one ghost mode, the Schwarzschild BH for the coupling range $|\alpha| \gg r_h^2$ is subject to Laplacian instabilities of several dynamical perturbations around the horizon in both along the radial and angular directions. In particular, the time scales of instabilities of large radial and angular momentum modes are typically much shorter than $r_h/c$, where $c$ is the speed of light. 
These Laplacian instabilities destabilize the Schwarzschild BH 
much more rapidly in comparison to the long-wavelength instability 
mentioned above.

This paper is organized as follows. In Sec.~\ref{scasec}, we revisit the SSS BH solutions present in Weyl gravity. In Sec.~\ref{oddsec}, we will discuss the propagation of dynamical perturbations in the odd-parity sector and show that all the speeds of propagation are luminal without classical Laplacian instabilities. In Sec.~\ref{evensec}, we will present the prescription for extracting four dynamical perturbations from the second-order action in the even-parity sector. We then show that  Laplacian instabilities emerge for both large radial and angular modes in the coupling range $|\alpha| \gg r_h^2$. Sec.~\ref{consec} is devoted to conclusions. Throughout the paper, we will use the natural units where 
the speed of light $c$ and the reduced Planck constant $\hbar$ are 1.

%%%%%%%%%%%%%%%%%%%%%%%%%%%%%%%%%%%%%%%%%%
\section{Black holes in quadratic Weyl gravity}
\label{scasec}
%%%%%%%%%%%%%%%%%%%%%%%%%%%%%%%%%%%%%%%%%%

We study the linear stability of BHs given by the action 
\be
{\cal S}=\frac{M_{\rm pl}^2}{2} 
\int {\rm d}^4 x \sqrt{-g} 
\left( R -\alpha C_{\mu \nu \rho \sigma} 
C^{\mu \nu \rho \sigma} \right)\,,
\label{Saction}
\ee
where $g$ is a determinant of the metric tensor $g_{\mu \nu}$. The Weyl squared term $C_{\mu \nu \rho \sigma} C^{\mu \nu \rho \sigma}$, which is given by Eq.~(\ref{Weyl}), is equivalent to 
$2R_{\mu \nu} R^{\mu \nu}-(2/3)R^2$ up to boundary terms. 

Taking into account the quadratic Ricci scalar $\beta R^2$ allows for the possibility of constructing renormalizable theories of gravity \cite{Stelle:1976gc}. 
However, the Weyl term gives rise to ghost DOFs, which violate the unitarity of theories. The ghosts arise from derivative terms higher than 
second order in the field equations of motion. Although the existence of ghosts in higher-derivative theories can be problematic, there are some arguments stating that the ghost can be ``soft'' in the sense that small classical perturbations are not subject to instabilities \cite{Smilga:2004cy}. 
In this paper, we would like to study the number of propagating DOFs on the SSS background and the linear stability of SSS BHs in Weyl gravity to see whether BHs are not prone to Laplacian instabilities. 
For this purpose, we focus on the effect of the Weyl curvature term without taking into account the $\beta R^2$ term. As in the analysis of Refs.~\cite{Lu:2015cqa, Lu:2015psa}, we do not deal with Weyl gravity as an effective field theory where the Weyl term $\alpha C_{\mu \nu \rho \sigma}C^{\mu \nu \rho \sigma}$ is suppressed relative to $R$.

We consider the SSS background given by the line element 
\be 
\rd s^2=-f(r) \rd t^{2} +h^{-1}(r) 
\rd r^{2}+ r^{2} \left( \rd \theta^{2}
+\sin^{2}\theta\,\rd\varphi^{2} 
\right)\,,
\label{background}
\ee
where $f$ and $h$ are functions of the radial coordinate $r$. We compute the action (\ref{Saction}) on the background (\ref{background}) and vary it with respect to $f$ and $h$. This gives fourth-order and third-order differential equations for $f$, with respect to the radial derivatives. Since the latter contains derivatives of $h$ up to the second order, we take the $r$ derivative of it and eliminate $f''''$ by using the former equation, where a prime represents the derivative with respect to $r$. After this procedure, we obtain a third-order differential equation for $f$. Combining it with the equation derived by the variation of $h$ leads to a second-order differential equation for $f$, as 
\be
f''=\frac{r^2 hf'^2-4(rh'+h-1)f^2
-(rh'+4h)r ff'}{2r^2 f h}\,,
\label{ddfeq}
\ee
which does not contain the Weyl coupling constant $\alpha$. 
We differentiate Eq.~(\ref{ddfeq}) with respect to $r$ and eliminate $f'''$ by exploiting the other third-order differential equation of $f$. Then, we obtain
\be
h''=\frac{(1-h)f-rf'h}{\alpha h(2f-rf')}
+\frac{4f^3 (h-1)(rh'+2h)+r^3f'^2 h (fh'-hf')
+r^2 f(3h^2 f'^2+2fh f' h'+3f^2 h'^2)}{2r^2 f^2 h (2f-rf')}.
\label{ddheq}
\ee
{}From Eqs.~(\ref{ddfeq}) and (\ref{ddheq}), we find that the Schwarzschild metric components
\be
f=h=1-\frac{r_h}{r}\,,
\label{Sch}
\ee
are the exact solution to the system, where $r_h$ is the horizon radius.
This Schwarzschild branch is present for any arbitrary coupling $\alpha~(\neq 0)$.

In Refs.~\cite{Lu:2015cqa, Lu:2015psa}, the authors numerically found the other non-Schwarzschild branch of BH solutions for positive $\alpha$ of order $r_h^2$. Although there are no exact solutions for this branch, it is possible to obtain an approximate solution by using a continued-fraction expansion of the non-GR solution \cite{Rezzolla:2014mua}. 
The metric components of the non-Schwarzschild branch can be 
expressed in the form 
\ba
f(r) \simeq \left( 1-\frac{r_h}{r} \right)A(r)\,,\qquad
h(r) \simeq \left( 1-\frac{r_h}{r} \right)\frac{A(r)}{B^2(r)}\,,
\label{nS1}
\ea
where $A(r)$ and $B(r)$ are functions of $r$ whose approximate 
formulas are given in Ref.~\cite{Kokkotas:2017zwt}. 
This hairy BH solution has a horizon at $r=r_h$ and 
it also respects the asymptotic flatness. 
In Ref.~\cite{Lu:2015cqa}, it was numerically shown that this non-Schwarzschild branch is present in the range
\be
0.876 \le p \le 1.143\,,
\label{prange}
\ee
where 
\be
p \equiv \frac{r_h}{\sqrt{2\alpha}}\,.
\ee
For $p>1.143$, the mass of non-Schwarzschild BHs becomes negative.
At the point $p=0.876$, the non-Schwarzschild and Schwarzschild branches intersect with each other. 
The non-Schwarzschild branch can be extended to the region $p<0.876$ down to the value $p \approx 0.6$, but it was shown in Ref.~\cite{Held:2022abx} that the solution in the range $p<0.876$ is prone to the long-wavelength instability related to the Gregory-Laflamme instability \cite{Gregory:1993vy}. To avoid this instability for the non-Schwarzschild branch, the variable $p$ needs to be in the range (\ref{prange}).

As we mentioned above, the Schwarzschild branch (\ref{Sch}) is present for any values of $\alpha$. The mass squared arising from the Weyl curvature computed on the Minkowski background is given by 
\be
m_W^2=\frac{1}{2\alpha}\,,
\label{mW}
\ee
which is positive if $\alpha>0$. For $p \gg 1$, i.e., 
in the ``massive'' regime characterized by $m_W \gg r_h^{-1}$, 
the Weyl term $\alpha C_{\mu \nu \rho \sigma}C^{\mu \nu \rho \sigma}$ is suppressed relative to $R$ outside the horizon. The other limit $p \ll 1$, which corresponds to the ``massless'' regime characterized by 
$m_W \ll r_h^{-1}$, the Weyl term dominates over the Ricci scalar. 
The latter is the regime in which the modification of gravity 
manifests itself in BH physics.

Since there is only the Schwarzschild branch for the mass range $m_W \ll r_h^{-1}$, we will exploit the background metric components (\ref{Sch}) to study the propagation of dynamical perturbations and linear stability of BHs for high radial and angular momentum modes. We note that the 
long-wavelength instability of Schwarzschild BHs is known to be present for $p<0.876$ \cite{Held:2022abx}. 
This long-wavelength mode is characterized by the 
monopole ($l=0$) with a negligible radial wavenumber 
relative to mass terms.
Since some of the dynamical perturbations vanish for $l=0$, we would like to clarify how all the dynamical perturbations propagate for the 
short-wavelength modes with $k r_h \gg 1$ and $l \gg 1$.

%%%%%%%%%%%%%%%%%%%%%%%%%%%%%%%%%%
\section{Odd-parity perturbations}
\label{oddsec}
%%%%%%%%%%%%%%%%%%%%%%%%%%%%%%%%%%

On the SSS background (\ref{background}) with the metric tensor $\bar{g}_{\mu \nu}$, we will study the linear stability of BHs 
in the presence of metric perturbations $h_{\mu \nu}$. Namely, the metric tensor of a perturbed line element is given by $g_{\mu \nu}=\bar{g}_{\mu \nu}+h_{\mu \nu}$. We focus on the BH stability in the external region of the horizon, i.e., $f(r)>0$ and $h(r)>0$.

We first consider $h_{\mu \nu}$ in the odd-parity sector where the perturbations have a parity $(-1)^{l+1}$ under the rotation in the $(\theta, \varphi)$ plane, with $l$ being the multipole of the spherical harmonics $Y_{l m} (\theta, \varphi)$. In the odd-parity sector, the components of $h_{\mu \nu}$ 
are given by \cite{Regge:1957td, Zerilli:1970se, DeFelice:2011ka, Motohashi:2011pw, Kobayashi:2012kh, Kase:2014baa}
\ba
& &
h_{tt}=h_{tr}=h_{rr}=0\,,\qquad 
h_{ab}=0\,,
\nonumber \\
& &
h_{ta}=\sum_{l,m}Q(t,r)E_{ab}
\nabla^bY_{lm}(\theta,\varphi)\,,
\qquad
h_{ra}=\sum_{l,m}W(t,r)E_{ab} 
\nabla^bY_{lm}(\theta,\varphi)\,,
\ea
where $Q$ and $W$ are functions of $t$ and $r$, the subscripts $a$ and $b$ denote either $\theta$ or $\varphi$, and $E_{ab}$ is an antisymmetric tensor with nonvanishing components $E_{\theta \varphi}=-E_{\varphi \theta}=\sin \theta$. Note that we have omitted the subscripts $l$ and $m$ in $Q$ and $W$ and chosen the gauge in which $h_{ab}$ vanishes.

Without loss of generality, we focus on the axisymmetric modes ($m=0$) 
and expand the action (\ref{Saction}) up to the second order in odd-parity perturbations. 
After the integration with respect to $\theta$ and $\varphi$, we can write the second-order action of perturbations in the form ${\cal S}_{\rm odd}^{(2)}=(\Mpl^2/2) \int {\rm d}t {\rm d}r\,\tilde{L}_{\rm odd}$, where $\tilde{L}_{\rm odd}$ is a function of $t$ and $r$ composed of the products of odd-parity perturbations.
In $\tilde{L}_{\rm odd}$, there exists the following combination
\be
(\tilde{L}_{\rm odd})_K 
\equiv -\frac{\alpha \Mpl^2 h^{1/2} l(l+1)}{2f^{3/2}}
\left( \ddot W -{\dot Q}' +\frac{2\dot{Q}}{r} 
\right)^2\,,
\label{LK}
\ee
where a dot represents the derivative with respect to $t$. This is equivalent to the following Lagrangian
\be
(L_{{\rm odd}})_K= -\frac{\alpha \Mpl^2 h^{1/2} l(l+1)}
{2f^{3/2}} \left[ 2\chi \left( \ddot{W} -{\dot Q}' 
+\frac{2\dot{Q}}{r} \right)-\chi^2
\right]\,,
\label{cLK}
\ee
where $\chi$ is a Lagrange multiplier. The Lagrangian $\tilde{L}_{\rm odd}-(\tilde{L}_{\rm odd})_K+(L_{\rm odd})_K$ is equivalent to the original one $\tilde{L}_{\rm odd}$. Varying the former with respect to $\chi$, it follows that 
\be
\chi=\ddot{W} -{\dot Q}' 
+\frac{2\dot{Q}}{r}\,,
\label{chidef}
\ee
which corresponds to a new dynamical DOF. 
After integrating the action 
$(\Mpl^2/2) \int {\rm d}t {\rm d}r\,
[\tilde{L}_{\rm odd}
-(\tilde{L}_{\rm odd})_K+(L_{\rm odd})_K]$
by parts, we can express the second-order action (up to boundary terms) in the form 
\be
{\cal S}^{(2)}_{\rm odd}=\frac{\Mpl^2}{2}
\int {\rm d}t {\rm d}r\,L_{\rm odd}\,,
\label{Sodd}
\ee
where
\ba
L_{\rm odd} &=&
a_1 \dot{W}^2+a_2 \dot{Q}^2
+2 a_3 \dot{W} \dot{\chi}
+a_4 \left( \dot{W}'-Q''+\frac{2Q'}{r} 
\right)^2
+a_5 W'^2+a_6 Q'^2
+a_7 W^2+a_8 \chi^2 +a_9 Q^2
\nonumber \\
& &
+a_{10} W' \dot{Q}
+a_{11} \dot{W} Q'+a_{12}\dot{\chi}Q'
+a_{13} \dot{W}Q
+a_{14} \chi \dot{Q}\,,
\label{Lag}
\ea
with $a_{1}, \cdots ,a_{14}$ being functions of $r$ alone. 
From this action, it is clear that there are three dynamical 
perturbations $W$, $Q$, and $\chi$.
The perturbation equations of motion follow by varying $L_{\rm odd}$ with respect to those variables.
We study the propagation of short-wavelength modes with the large angular frequency $\omega$ and momentum $k$ by assuming the solutions in the form
\be
\vec{\cal X}=\vec{\cal X}_0 
e^{i (\omega t-kr)}\,,\qquad 
{\rm with} \qquad 
\vec{\cal X}_0=(W_0, \chi_0, Q_0)\,,
\label{cX}
\ee
where $W_0$, $\chi_0$, and $Q_0$ are assumed to be constants.
We are interested in the values of $k$ and $l$ in the ranges $k r_h \gg 1$ and $l \gg 1$. Note that we also focus on the WKB regime in which the radial variation of $\omega$ is small such that $|\omega'|\ll |k\omega|\simeq|\omega^2|$. 

In the limit that $l \gg 1$, each coefficient in Eq.~(\ref{Lag}) 
has the following multipole dependence:
\ba
& &
a_1=b_1 l^4\,,\quad
a_2=b_2 l^4\,,\quad
a_3=b_3 l^2\,,\quad
a_4=b_4 l^2\,,\quad
a_5=b_5 l^4\,,\quad
a_6=b_6 l^4\,,\quad
a_7=b_7 l^6\,,\quad
a_8=b_8 l^2\,,\nonumber \\
& &
a_9=b_9 l^6\,,\quad
a_{10}=b_{10} l^4\,,\quad
a_{11}=b_{11} l^2\,,\quad
a_{12}=b_{12} l^2\,,\quad
a_{13}=b_{13} l^4\,,\quad
a_{14}=b_{14} l^2\,,
\ea
where 
\ba
& &
b_1=\frac{2\alpha h^{1/2}}{r^2 f^{1/2}}\,,\qquad 
b_2=-\frac{1}{2fh}b_1\,,\qquad
b_3=\frac{r^2}{2f}b_1\,,\qquad 
b_4=\frac{r^2 h}{2}b_1\,,\qquad 
b_5=-\frac{fh}{2}b_1\,,\qquad 
b_6=b_1\,,\nonumber \\
& &
b_7=-\frac{f}{2r^2}b_1\,,\qquad 
b_8=\frac{r^2}{2f}b_1\,,\qquad 
b_9=\frac{1}{2r^2 h}b_1\,,\qquad 
b_{10}=-b_1\,,\qquad 
b_{12}=-\frac{r^2}{f}b_1\,.
\ea
The explicit forms of $b_{11}$, $b_{13}$, and $b_{14}$ are not shown, as they are not needed in the following discussion.

Picking up the dominant contributions of $\omega$, $k$ and $l$, the perturbation equations of motion are expressed as
\be
{\bm A}_{\rm odd}
\vec{{\cal X}}_0^{\,\rm T}=0\,,
\label{Aeq}
\ee
where ${\bm A}_{\rm odd}$ is a $3 \times 3$ matrix 
whose components are given by 
\be
{\bm A}_{\rm odd}=\left(\begin{array}{ccc}
2l^2 \left[(b_4 k^2+b_1 l^2)\omega^2 
+ b_5 k^2 l^2 + b_7 l^4 \right] & 2l^2 b_3 \omega^2 & 
l^2(2b_4k^3 \omega -b_{10} l^2k \omega) \\
2l^2 b_3 \omega^2 & 2l^2 b_8 & -l^2 b_{12}k\omega\\
l^2(2b_4k^3 \omega -b_{10} l^2k \omega) & 
-l^2 b_{12}k\,\omega 
& 2 l^2 \left(b_2 l^2 \omega^2 +b_4 k^4 
+ b_6 k^2 l^2 + b_9 l^4 \right)
\end{array}\right)\,.
\label{Aodd}
\ee

The no-ghost conditions can be obtained by picking up terms proportional to $\omega^2$ in ${\bm A}_{\rm odd}$. The matrix ${\bm K}_{\rm odd}$ associated with such kinetic terms is 
\be
{\bm K}_{\rm odd}=2\omega^2 \left(\begin{array}{ccc}
{\cal K}_{11} & {\cal K}_{12} & 0\\
{\cal K}_{12} & 0 & 0\\
0 & 0 & {\cal K}_{33}
\end{array}\right)\,,
\ee
where 
\be
{\cal K}_{11}=l^2 (b_4 k^2+b_1 l^2)\,,\qquad 
{\cal K}_{12}=l^2 b_3\,,\qquad 
{\cal K}_{33}=l^4 b_2\,.
\ee
The absence of ghosts requires the following three conditions
\be
{\cal K}_{11}>0\,,\qquad -{\cal K}_{12}^2>0\,,
\qquad 
-{\cal K}_{12}^2 {\cal K}_{33}>0\,.
\label{noghost}
\ee
The explicit form of ${\cal K}_{12}$ is given by 
\be
{\cal K}_{12}=\frac{\alpha h^{1/2} l^2}{f^{3/2}}\,.
\ee
Since the second inequality of (\ref{noghost}) is violated for $\alpha \neq 0$, there is at least one ghost mode in the odd-parity sector. 
After making the field redefinitions $W=W_2-{\cal K}_{12}\,{\chi_2}/{\cal K}_{11}$, $\chi=\chi_2$, and $Q=r Q_2$, the kinetic matrix 
of new fields $(W_2, \chi_2, Q_2)$ becomes diagonal with 
the elements ${\cal K}_{11}$, 
$-{\cal K}_{12}^2/{\cal K}_{11}$, 
and ${\cal K}_{33}$, where 
\be
{\cal K}_{11}=\frac{\alpha l^2 \sqrt{h}(r^2 h k^2+2l^2)}
{r^2 \sqrt{f}}\,,\qquad
-\frac{{\cal K}_{12}^2}{{\cal K}_{11}}
=-\frac{\alpha l^2 r^2 \sqrt{h}}
{f^{5/2} (r^2 h k^2+2l^2)}\,,\qquad
{\cal K}_{33}=-\frac{\alpha l^4}{f^{3/2} h^{1/2}} \,.
\label{Koddcom}
\ee
For $\alpha>0$, there are two ghosts because the last two eigenvalues are negative. For $\alpha<0$, one ghost is present because $K_{11}$ is negative. It should be noticed that all the elements of Eq.~(\ref{Koddcom}), for large values of $r$, do not vanish but tend to approach constant values. This is related to the fact that the mass squared for these modes, in the regime $r \gg r_h$, is of the same order as $m_W^2=1/(2\alpha)$, i.e., a finite value independent of the radial distance $r$. 
From this analysis of the kinetic matrix, we can conclude that the odd-parity perturbations in this theory do not suffer from a strong coupling problem. This is in contrast with the odd-parity perturbations in Einsteinian cubic gravity, in which the strong coupling problem 
is present \cite{DeFelice:2023vmj, Jimenez:2023esa}.

For the radial propagation, we calculate the determinant of the matrix ${\bm A}_{\rm odd}$ and expand it with respect to the large momentum $k$ in the regime $k r_h \gg l \gg 1$. The radial propagation speeds $c_r={\rm d}r_*/{\rm d}\tau$, which are measured by the rescaled radial coordinate $r_*=\int {\rm d}r/\sqrt{h}$ and the proper time $\tau=\int \sqrt{f}\,{\rm d}t$, are given by $c_r=(fh)^{-1/2}(\partial\omega/\partial k)$. The dominant terms in the equation ${\rm det}\,{\bm A}_{\rm odd}=0$ are those proportional to $k^6$, so that we obtain the following three solutions 
\ba
c_{r1}^2 &=&
\frac{b_4 b_8 f^2}{\alpha^2 h^2}=1\,,\label{cr1}\\
c_{r2}^2 &=&
\frac{-(b_1+b_6+b_{10})
+\sqrt{{\cal D}_1}}{2b_2 fh}=1\,,\\
c_{r3}^2 &=&
\frac{-(b_1+b_6+b_{10})
-\sqrt{{\cal D}_1}}{2b_2 fh}
=1\,,
\label{cr3}
\ea
where
\ba
{\cal D}_1=b_1^2+2b_1 b_6-4 b_2 b_5+b_6^2
+2(b_1+b_6)b_{10}+b_{10}^2=0\,.
\ea
Thus, all three radial propagation speeds are luminal. 

For the angular propagation, we expand ${\rm det}\,{\bm A}_{\rm odd}$ 
with respect to large $l$ in the range $l \gg k r_h \gg 1$.
The angular propagation speeds $c_{\Omega} =r {\rm d}\theta/{\rm d}\tau$ measured by the proper time are $c_{\Omega}=r\omega/(\sqrt{f} l)$. 
The leading-order terms, which are proportional to $l^{14}$, 
give rise to the following three solutions
\ba
c_{\Omega 1}^2 &=& 
\frac{r^2(b_1 b_8+\sqrt{{\cal D}_2})}{2b_3^2 f}=1\,,
\label{cO1}\\
c_{\Omega 2}^2 &=& 
\frac{r^2(b_1 b_8-\sqrt{{\cal D}_2})}{2b_3^2 f}
=1\,,\label{cO2}\\
c_{\Omega 3}^2 &=& -\frac{b_9 r^2}{b_2 f}
=1\,,
\label{cO3}
\ea
where
\ba
\hspace{-0.3cm}
{\cal D}_2 &=&
b_8 (b_1^2 b_8+4 b_3^2 b_7)=0
\,.
\ea
Hence all three angular propagation speeds are also luminal.

Since we have not used the background solutions of $f(r)$ and $h(r)$ to derive the above propagation speeds, they are valid for any BH solutions under the scheme of the WKB approximation. 
In this same short-wavelength regime, this analysis shows 
that ghost modes present in the odd-parity sector are ``soft.'' 
We have thus shown that there is at least one ghost mode in the odd-parity sector, but the classical Laplacian instabilities are absent for both radial and angular directions. 

%%%%%%%%%%%%%%%%%%%%%%%%%%%%%%%%%%
\section{Even-parity perturbations}
\label{evensec}
%%%%%%%%%%%%%%%%%%%%%%%%%%%%%%%%%%

For the even-parity sector, we consider the components of metric perturbations $h_{\mu \nu}$ as 
\begin{alignat}{6}
h_{tt}&=f(r)\sum_{l,m}H_0(t,r)Y_{lm}(\theta,\varphi)\,,\qquad & 
h_{tr}&=\sum_{l,m}H_1(t,r)Y_{lm}(\theta,\varphi)\,, \qquad
&h_{ta}&=0\,,\nonumber\\
h_{rr} &= h(r)^{-1}\sum_{l,m}H_2(t,r)Y_{lm}(\theta,\varphi)\,,\qquad &h_{ra}&=\sum_{l,m}h_1(t,r)\nabla_a Y_{lm}(\theta,\varphi)\,,\qquad&h_{ab}&=0\,,
\end{alignat}
where $H_0$, $H_1$, $H_2$, and $h_1$ depend on $t$ and $r$. 
We have chosen the gauge conditions $h_{ta}=0=h_{ab}$, 
which fix the residual gauge DOFs.

We expand the action up to quadratic order in even-parity perturbations by setting $m=0$. After the integration with respect to $\theta$ and $\varphi$, the second-order action can be expressed in the form 
\be
{\cal S}^{(2)}_{\rm even}
=\int {\rm d}t {\rm d}r\,\tilde{L}_{\rm even}\,,
\ee
where $\tilde{L}_{\rm even}$ is the Lagrangian containing the products of even-parity perturbations. 
First of all, we notice that there are higher-order time derivative terms $\ddot{H}_{2}^{2}$ and $\ddot{h}_{1}^{2}$ in $\tilde{L}_{\rm even}$. 
To find the combinations of Lagrange multipliers $\chi_1$ and $\chi_2$ associated with $\ddot{H}_{2}$ and $\ddot{h}_{1}$, respectively, we consider the following Lagrangian 
\ba
L_{\rm even} &=& 
\tilde{L}_{\rm even}+\frac{\alpha \Mpl^{2} r^{2}}{6h^{1/2}f^{3/2}}\left(\ddot{H}_{2}
+c_{1}H''_{0}+c_{2}\dot{H}'_{1}+c_{3}H'_{0}
+c_{4}\dot{H}_{1}+c_{5}H_{0}+c_{6}H_{1}+c_{7}H'_{1}-\chi_{1}\right)^{2}\nonumber \\
&& +\frac{l(l+1)\alpha\Mpl^{2}h^{1/2}}{2f^{3/2}}\left(\ddot{h}_{1}+d_{1}H''_{0}+d_{2}\dot{H}'_{1}
+d_{3}H'_{0}+d_{4}\dot{H}_{1}+d_{5}H_{0}+d_{6}H_{1}
+d_{7}H'_{1}-\chi_{2}\right)^{2}\,.
\label{Leven}
\ea
We need to choose the $r$-dependent coefficients $c_i$, $d_i$ ($i=1,2,\cdots, 7$) to eliminate the cross products such as $\ddot{H}_2H_0''$ and $\ddot{h}_1 H_0'$. Then, these coefficients are determined as  
\ba
& &
c_1=fh\,,\qquad c_2=-2h\,,\qquad 
c_3=hf'+\frac{1}{2}fh'-\frac{fh}{r}\,,\qquad
c_4=\frac{2h}{r}-h'\,,\nonumber \\
& &
c_5=\frac{l(l+1)f}{2r^{2}}\,,\qquad 
c_6=0\,,\qquad
c_7=0\,,\label{cidef} \\
& &
d_1=0\,,\qquad d_2=0\,,\qquad 
d_3=f\,,\qquad d_4=-1\,,\qquad
d_5=\frac{1}{2}\,f'-\frac{f}{r}\,,\qquad
d_6=0\,,\qquad d_7=0\,.
\label{didef}
\ea
Varying $L_{\rm even}$ with respect to $\chi_1$ and $\chi_2$, respectively, it follows that
\ba
\chi_1 &=& \ddot{H}_2+fh H_0''-2h \dot{H}_1'
+\left( hf'+\frac{1}{2}fh'-\frac{fh}{r} \right)H_0'
+\left( \frac{2h}{r}-h' \right) \dot{H}_1
+\frac{l(l+1)f}{2r^{2}}H_0\,,\\
\chi_2 &=& \ddot{h}_1+f H_0'-\dot{H}_1
+\left( \frac{1}{2}\,f'-\frac{f}{r} \right)H_0\,,
\ea
both of which correspond to the propagating DOFs.
As we will see below, there are four propagating dynamical 
perturbations in the even-parity sector (including $\chi_1$ and $\chi_2$).

The Lagrangian $L_{\rm even}$ contains the following quadratic term 
\be
L_{\rm even} \ni-\frac{\alpha \Mpl^{2} 
l (l+1)[l(l+1)-2]
\sqrt{f}}{8r^{2}\sqrt{h}}\,H_{0}^{2}\,,
\ee
besides the linear terms in $H_0$. 
Hence we will integrate the nondynamical perturbation $H_0$ from 
the second-order action.

The term proportional to $\dot{H}_{1}^{2}$ disappears in $L_{\rm even}$, because of the choice of coefficients $c_{4}$ and $d_{4}$ in Eqs.~(\ref{cidef}) and (\ref{didef}). 
The Lagrangian $L_{\rm even}$ does not contain the time derivatives of $H_1$, so the perturbation $H_1$ is not dynamical either.
For later convenience, we perform the following field redefinition 
\be
{\bar H}_1 \equiv H_1 f^{-1/4} h^{3/4}\,.
\label{barH1}
\ee
Then, we find that the ${\bar H}_1$-dependent terms can be expressed as
\be
L_{\rm even} \ni \frac12\alpha\Mpl^2l(l+1)({\bar H}_1')^2
+e_1 {\bar H}_1^2 + {\bar H}_1
\bigl( e_2{\dot h}_1''+e_3{\dot H}_2''+e_4{\dot h}_1'+e_5{\dot H}_2'+e_6{\dot\chi}_1'+
e_7{\dot h}_1+e_8{\dot H}_2+e_{9}{\dot \chi}_1
+e_{10}{\dot \chi}_2 \bigr),
\ee
where the coefficients $e_i$'s ($i={1\dots10}$) are $r$-dependent functions. 
It should be noticed that ${\bar H}_1$ does not appear 
anywhere else inside $L_{\rm even}$. 
It is then clear that ${\bar H}_1$ is not a propagating field that can be integrated out from the second-order action.
This should leave only four dynamical DOFs in the even-parity sector. 
At this moment, to disentangle as much as possible the dynamics of ${\bar H}_1$ with those of the other perturbations, we perform another field redefinition as 
\be
-\Mpl^2 l(l+1){\bar \chi}_2 \equiv e_2{h}_1''+e_3{H}_2''
+e_4{h}_1'+e_5{H}_2'+e_6{\chi}_1'+
e_7{h}_1+e_8{H}_2+e_{9}{\chi}_1
+e_{10}{\chi}_2\,.
\label{bchi2}
\ee
The new variable ${\bar \chi}_2$ is used instead of ${\chi}_2$ in the following discussion. 

At this point, we will proceed by performing several integrations by parts, as presented in Appendix \ref{app:by_parts}. 
This step is meant to bring the action in a canonical form just before integrating out the nondynamical field $\bar{H}_1$.
Then, up to boundary terms, we can express $L_{\rm even}$ in the following form
\begin{align}
    L_{\rm even}&=A_{ij}^{(1)}\,{\dot\psi}'_i\,{\dot\psi}'_j
    +A_{ij}^{(2)}\,{\dot\psi}_i\,{\dot\psi}_j
    +\frac12\,B_{ij}^{(1)}\,
    \left( {\dot\psi}'_i\,{\dot\psi}_j-{\dot\psi}'_j\,{\dot\psi}_i \right)+A_{ij}^{(3)}\,\psi_i'''\psi_j'''
    +\frac12\,B_{ij}^{(2)}\,\left(\psi'''_i\,\psi_j''-\psi'''_j\,\psi_i'' \right)
    \nonumber\\
    &+A_{ij}^{(4)}\,\psi_i''\psi_j''
    +\frac12\,B_{ij}^{(3)}\,
    \left(\psi''_i\,\psi_j'-\psi''_j\,\psi_i' \right)
    +A_{ij}^{(5)}\,\psi_i'\psi_j'
    +\tfrac12\,B_{ij}^{(4)}\,
    \left( \psi'_i\,\psi_j-\psi'_j\,\psi_i \right)
    +A_{ij}^{(6)}\,\psi_i\psi_j
    \nonumber\\
    &+\frac12\alpha\Mpl^2l(l+1)({\bar H}_1')^2
    +e_1 {\bar H}_1^2 -\Mpl^2 l(l+1) {\bar H}_1 \dot\psi_4\,,
    \label{eq:canonical_L}
\end{align}
where $A_{ij}^{(j)}$'s and $B_{ij}^{(j)}$'s are symmetric and antisymmetric matrices, respectively, and $\psi_i$'s consist of four dynamical perturbations given by 
\be
\psi_1=h_1\,,\qquad 
\psi_2=\chi_1\,,\qquad 
\psi_3=H_2\,,\qquad 
\psi_4={\bar\chi}_2\,.
\ee
In Eq.~(\ref{eq:canonical_L}), there is a radial derivative 
term $({\bar H}_1')^2$. As such, one is required to set boundary conditions on the nondynamical field ${\bar H}_1$, which may eventually influence the dynamics of other propagating fields, in a fashion similar to the one present in theories that possess shadowy modes \cite{DeFelice:2018ewo, DeFelice:2021hps}.

We need to take care when we integrate out the field ${\bar H}_1$ from the second-order action. 
To compute the speeds of propagation of dynamical perturbations, we need to assume the WKB approximation to hold. This approximation is valid so long as the coefficients of the differential equations are slowly varying functions of $r$.
If we take a closer look at the equation of motion for ${\bar H}_{1}$ and evaluate it on the Minkowski background for simplicity, we obtain the differential equation 
\be
\bar{H}_1''-\left[ m_W^2+\frac{l(l+1)}{r^2} \right] 
\bar{H}_1
+2m_W^2\dot{\psi}_4 \simeq 0\,,
\label{H1eq}
\ee
where $m_W^2$ is given by Eq.~(\ref{mW}).
Recall that $m_W^2$ corresponds to the mass squared arising from the Weyl curvature term. 

For $\alpha>0$, the general solution to Eq.~(\ref{H1eq}) can be expressed as 
\ba
\bar{H}_1
&=& c_1 \sqrt{r} I_{l+1/2} (m_W r)+
c_2\sqrt{r} I_{l+1/2} (m_W r) \nonumber \\
& &
+2m_W^{2}\sqrt{r} \left[ K_{l+1/2}(m_Wr)
\int \sqrt{r}I_{l+1/2}(m_W r)\dot{\psi}_4 {\rm d}r
-I_{l+1/2}(m_Wr)\int 
\sqrt{r}K_{l+1/2}(m_W r) \dot{\psi}_4 {\rm d}r 
\right]\,, 
\label{H1sol}
\ea
where $c_1$, $c_2$ are integration constants, and $I_{l+1/2}(x)$ and $K_{l+1/2}(x)$ are the modified Bessel functions of the first and second kinds, respectively. 
On using the growing-mode solution $I_{l+1/2}(m_Wr)$ in the regime $m_W r \gg 1$, the radial derivative of $\bar{H}_1$ can be estimated as 
$|r \bar{H}_1'/\bar{H}_1| \approx m_W r \gg 1$. 
We would like to consider the WKB regime in which the solution to $\bar{H}_1$ is expressed in the form ${\bar H}_1=\tilde{\bar{H}}_1 e^{-i (\omega t-kr)}$, where $\omega$ and $k$ are the constant angular frequency and momentum, respectively, and $\tilde{\bar{H}}_1$ is constant.
In this regime, the radial variation of ${\bar H}_1$ is dominated by  
large momentum modes $k$, such that $|r \bar{H}_1'/\bar{H}_1| 
\approx kr \gg 1$. 
Since we need to avoid the large radial variation of $\bar{H}_1$ induced by the mass term $m_W$, we require the condition $m_W r \ll 1$.

For $\alpha<0$, the solution to $\bar{H}_1$ can be expressed in terms of $J_{l+1/2}(\sqrt{-m_W^2}r)$ and $Y_{l+1/2}(\sqrt{-m_W^2}r)$, where $J_{l+1/2}(x)$ and $Y_{l+1/2}(x)$ are the Bessel functions of first and second kinds, respectively.
In this case the perturbation $\bar{H}_1$ exhibits fast oscillations with respect to $r$, so the validity of the WKB approximation requires the condition $\sqrt{-m_W^2} r \ll 1$.

The above discussion shows that for both $\alpha>0$ and $\alpha<0$, the condition $|m_W^2| r_h^2 \ll 1$ is needed to ensure the validity of the WKB approximation in the vicinity of the BH horizon $r_h$. 
This translates to the condition 
\be
|\alpha| \gg r_h^2\,.
\label{alrange}
\ee
In the following, we will consider the finite region of $r$ 
satisfying the inequality 
\be
r_h^2 \lesssim r^2 \ll |\alpha|\,.
\ee
Thus, our analysis based on the WKB approximation 
does not encompass the small coupling region 
$|\alpha| \lesssim r_h^2$ 
due to the dominance of a heavy mass squared $|m_W^2|$. 
In other words, the linear stability of even-parity perturbations on the Minkowski background with the coupling range $|\alpha| \ll r_h^2$ cannot be 
obtained by simply taking the limit $r \gg r_h$ for the BH stability conditions derived below. 

We should make it clear that the inequality $|\alpha| \gg r_h^2$ 
is not a condition for stability, but merely a condition under 
which the WKB approximation works. In other words, in that region 
of the parameter space, we can trust the approximate but analytic 
prescriptions presented below. 
For the parameter range (\ref{alrange}) there is only the 
Schwarzschild branch (\ref{Sch}), but the non-Schwarzschild
branch with the metric components (\ref{nS1}) 
is not present. Hence the BH linear stability discussed 
below can be only applied to the Schwarzschild solution 
(\ref{Sch}) in the coupling regime where the Weyl curvature 
term dominates over $R$.

If the parameter space in Weyl gravity lies outside the range 
(\ref{alrange}), then the theory might still be unstable, 
but a different analysis, either analytical or numerical, 
is necessary to establish the stability of the background BH solution. 
For instance, with a largely negative value of the squared mass 
of perturbations, namely for $-M^2 < 1/\alpha<0$ (where $M$ is 
a cutoff mass scale of the theory), one could expect 
tachyonic instability to occur.

For what we have discussed so far, we will not describe how the solutions depend on the choice of boundary conditions for ${\bar H}_1$ at spatial infinity. Instead, we will only consider some subset of solutions for the theory, which can be treated under the WKB approximation. 
Then, we assume the solutions to $\psi_i$'s and ${\bar H}_1$ in the forms 
\be
\psi_i=\tilde{\psi}_i e^{-i (\omega t-kr)}\,,\qquad 
{\bar H}_1=\tilde{\bar{H}}_1 e^{-i (\omega t-kr)}\,,
\ee
where $\tilde{\psi}_1=\tilde{h}_1$, 
$\tilde{\psi}_2=\tilde{\chi}_1$,
$\tilde{\psi}_3=\tilde{H}_2$,  
$\tilde{\psi}_4=\tilde{\bar{\chi}}_2$ 
and $\tilde{\bar{H}}_1$ are assumed to be constant.
Although we focus only on these solutions, if Laplacian instabilities 
occur for them, then the SSS background becomes also unstable in general. 
Under the WKB approximation, we turn our attention to the large $k$ and $l$ modes.
Furthermore, for any $r$-dependent coefficient ${\cal F}(r)$, we will exploit the approximation $({\cal F}(r)\,\psi_i)' \approx {\cal F}(r)\,\psi_i' \to ik\,{\cal F}(r)\,\tilde{\psi}_i$. 
In other words, when we vary the Lagrangian \eqref{eq:canonical_L}, the 
components of symmetric matrices $A_{ij}^{(j)}$'s and antisymmetric ones $B_{ij}^{(j)}$'s can be treated as constants. 

After integrating out $\tilde{\bar{H}}_1$, the field equations of motion for the four dynamical perturbations $\vec{\psi}=(\tilde{h}_1,
\tilde{\chi}_1,\tilde{H}_2,\tilde{\bar{\chi}}_2)$ can be expressed in the form 
\be
{\bm A}_{\rm even} \vec{\psi}^{\,\rm T}=0\,,
\label{Aeveneq}
\ee
where ${\bm A}_{\rm even}$ is a $4 \times 4$ Hermitian matrix whose components satisfy 
$({\bm A}_{\rm even})_{ij}=({\bm A}_{\rm even})^*_{ji}$.
We note that $({\bm A}_{\rm even})_{ij}$'s contain $r$-dependent background functions besides $k$ and $l$.

\subsection{No-ghost conditions}

The no-ghost conditions can be derived by considering terms proportional to $\omega^2$ in ${\bm A}_{\rm even}$. 
We express the matrix in ${\bm A}_{\rm even}$ containing the $\omega^2$ dependence as ${\bm K}_{\rm even}=\omega^2 {\bm K}$, where ${\bm K}$ is a $4 \times 4$ matrix whose components are given by $K_{ij}$.
To avoid ghosts in the even-parity sector, the determinants of submatrices of ${\bm K}$ need to be positive. 
In the limit of large values of $k$ and $l$, the no-ghost conditions translate to
\begin{align}
g_4&=K_{44} \simeq 
-\frac{\Mpl^2 \,r^{2} h\,l^{2}}
{\alpha \left(k^{2}r^{2}h+l^{2} \right)}>0
\,,\label{NG1} \\
g_3&=\frac{K_{33}K_{44}-K_{34}K_{43}}{g_4}
\simeq 
\frac{2 \alpha \Mpl^2 l^{2}}{3 \sqrt{fh}}>0
\,,\label{NG2} \\
g_2&=\frac{1}{g_3\,g_4}
\det
\left|\begin{array}{ccc}
K_{22} &K_{23} &K_{24}\\
K_{32} & K_{33} & K_{34}\\
K_{42} & K_{43} & K_{44}
\end{array}\right|
\simeq 
-\frac{\alpha \,r^{4} \Mpl^2}
{6 f^{5/2} \sqrt{h}\,l^{2}}>0\,,\label{NG3}\\
g_1&=\frac{\det({\bm K})}{g_2\,g_3\,g_4}\simeq 
-\frac{7\Mpl^2 (rh f'+fh-f)l^2}{12\sqrt{fh}(rf'-2f)}\nonumber\\
&~~~+\frac{\alpha\Mpl^2}{3 f^{5/2}\sqrt{h} r^{2} 
\left(r f'-2 f \right)}\Biggl\{ 
\left[ 4 \left(2 k^{2} r^{2}-2 k^{2} h' r^{3}-l^{2}
\right) h^{2}-l^{2} \left( \frac{15 r h'}{2}+8\right) h 
+\frac{r h' \left( 19 r h'-14 \right) l^{2}}{4}\right] f^{3}
\nonumber\\
&~~~+2 \left[ \left(k^{2} h' r^{3}-2 k^{2} r^{2}+\frac{61}{4} l^{2}\right) h^{2}-8 h^{3} k^{2} r^{2}+\frac{l^{2} \left( 23r h'
+2 \right) h}{8}-\frac{17 {h'}^{2} l^{2} r^{2}}{32}\right] 
r f' f^{2}\nonumber\\
&~~~+h {f'}^{2} r^{2} \left[ 
10 h^{2} k^{2} r^{2}+\left(k^{2} h' r^{3}-15 l^{2}\right) h 
+\frac{13 r h' l^{2}}{8}\right] f -h^{2} {f'}^{3} \left(h \,k^{2} r^{2}-\frac{27 l^{2}}{16}\right) r^{3} \Biggr\}>0\,,
\label{NG4}
\end{align}
where $g_i$'s correspond to the eigenvalues of ${\bm K}$. 
To derive the results (\ref{NG1})-(\ref{NG4}), we have not specified the background to be the Schwarzschild solution (\ref{Sch}).

Irrespective of the signs of $\alpha$, either of the three conditions (\ref{NG1})$\sim$(\ref{NG3}) is violated and hence there is always at least one ghost mode in the even-parity sector. The fourth condition (\ref{NG4}) is complicated, but if we evaluate it on the Schwarzschild solution (\ref{Sch}), we find
\be
g_1=\left[\frac{\left(8 \beta^{2}-20 \beta +15\right) l^{2}}
{4 \beta^{3}(\beta -1) r_{h}^{2}}
-\frac{4 (\beta-1)(\beta-3)k^{2}}{3 \beta^{2}}\right] 
\alpha \Mpl^2 \,,
\ee
where 
\be
\beta \equiv \frac{r}{r_h}\,.
\ee
In the vicinity of the horizon ($\beta=1$), we have the approximate relation $g_1 \simeq 3\alpha \Mpl^2 l^2/[4r_h^2(\beta-1)]$ and hence $g_1>0$ for $\alpha>0$ and $\beta>1$.

\subsection{Radial speeds of propagation}

To have nonvanishing solutions for $\vec{\psi}=(\tilde{h}_1,
\tilde{\chi}_1,\tilde{H}_2,\tilde{\bar{\chi}}_2)$, we require that Eq.~(\ref{Aeveneq}) obeys the discriminant equation
\be
{\rm det}\,{\bm A}_{\rm even}=0\,.
\label{detAeq}
\ee
The propagation speeds $c_r$ along the radial direction can be derived by using the approximation of large values of $\omega$ and $k$ in the range $k r_h \gg l \gg 1$.
Then, Eq.~(\ref{detAeq}) is approximated as
\be 
\mu_1k^2\omega^8
+\mu_2k^{6}\omega^6
+\mu_3k^{10}\omega^4
+\mu_4k^{14}\omega^2
+\mu_5k^{16}\simeq 0\,,
\label{eq:rad_discriminant}
\ee
where we picked up the dominant contribution in $k$ for each single $\omega^p$ coefficient with $p=\{0,2,4,6,8\}$, so that $\mu_i$'s depend on only $r$ and $l$. 
In general, we do not have the solution $\omega=0$ to Eq.~(\ref{eq:rad_discriminant}).
To find approximate dispersion relations which are valid for large values of $\omega$ and $k$ (with $k r_h\gg l \gg 1$), we will proceed as follows. 
First, we search for solutions of the kind $\omega=\sqrt{{\cal B}}\,k \propto k$. 
Substituting this expression into Eq.~\eqref{eq:rad_discriminant}, it follows that 
\be
k^{16}\,[\mu_4 {\cal B}+\mu_5+\mathcal{O}(k^{-2})]
\simeq 0\,,
\ee
and hence we obtain
\be
{\cal B}=-\frac{\mu_5}{\mu_4} \simeq \sqrt{fh}\,,
\ee
where we have finally taken the limit $l \gg 1$ for the computation of ${\cal B}$. 
Then, one of the dynamical perturbations has the dispersion relation 
\be
\omega=\sqrt{fh}\,k\,,
\ee
so that the corresponding radial propagation speed $c_{r1, {\rm even}}=(fh)^{-1/2}(\partial\omega/\partial k)$ is given by
\be
c_{r1, {\rm even}}=1\,,
\ee
which is luminal. On the other hand, we look for solutions of the kind $\omega=\sqrt{{\cal C}}\,k^2$. 
In this case, Eq.~(\ref{eq:rad_discriminant}) yields
\be
{\cal C} k^{18}\,[\mu_1 {\cal C}^3
+\mu_2 {\cal C}^2+\mu_3 {\cal C}
+\mu_4+\mathcal{O}(k^{-2})] \simeq 0\,,
\label{Ceq}
\ee
which leads to three non-trivial solutions for ${\cal C}$. Then, we already have a set of four required dispersion relations corresponding to dynamical perturbations.
Indeed, another ansatz, such as $\omega \propto k^3$ would not lead to any nontrivial solutions. 
For the Schwarzschild solution (\ref{Sch}) with $l \gg 1$, the coefficients are approximately given by 
\begin{align}
\mu_1&=l^8(\beta-3)\,,\qquad 
\mu_2=-\frac{l^{6} r_{h}^{2} \left(\beta -1\right)^{2} \left(4 \beta^{2}-34 \beta +39\right)}{6 \beta }\,,\nonumber\\ 
\mu_3&=\frac{l^{2} r_{h}^{4}
\left(352 \beta^{4}-2792 \beta^{3}+7512 \beta^{2}-8100 \beta +2997\right)\left(\beta -1\right)^{4}}{36 \beta^{3} }\,,\nonumber\\
\mu_4&=-\frac{r_{h}^{6} \left(\beta -1\right)^{7} \left(28 \beta^{2}-122 \beta +105\right)^{2}}{108 \beta^{4}}\,.
\end{align}
In general, the solution of Eq.~(\ref{Ceq}) depends on the value of $\beta=r/r_h$. 
Just to give an example, let us study the stability of the Schwarzschild BH at $\beta=2$. 
In this case, we obtain the three solutions ${\cal C}_1=13r_h^2/(12l^2)$, ${\cal C}_{2,3}=\pm 9\sqrt{13}r_h^2/(52l^3)$, and hence the corresponding propagation speeds are 
\ba
c_{r2,{\rm even}}&=& \frac{2\sqrt{39}}{3}\,\frac{kr_h}{l}\,,
\label{ome2}\\
c_{r3,{\rm even}} &=& \frac{6}{13^{1/4}}\,\frac{kr_h}{l^{3/2}}\,,
\label{ome3}\\
c_{r4,{\rm even}}^2 &=& -\frac{36}{\sqrt{13}}\,\frac{k^2r_h^2}{l^{3}}\,.
\label{ome4}
\ea
This shows that two dynamical perturbations corresponding to $c_{r2,{\rm even}}$ and $c_{r3,{\rm even}}$ are superluminal for $kr_h \gg l \gg 1$. 
Since $c_{r4,{\rm even}}^2$ is negative, there is Laplacian instability for one of the dynamical perturbations.
In the above discussion, we considered the case $\beta=r/r_h=2$, but the appearance of Laplacian instability for a distance close to the horizon is sufficient to exclude the Schwarzschild BH with the Weyl coupling constant $\alpha$ in the range (\ref{alrange}).

On using the propagation speed squared (\ref{ome4}), the time scale of Laplacian instability can be estimated as
\be
t_{\rm ins} \approx \frac{1}{\sqrt{l}} 
\left( \frac{l}{k r_h} \right)^2 r_h 
\ll r_h\,.
\ee
For the Schwarzschild BH with the horizon size $r_h \approx 10$~km, 
we have $t_{\rm ins} \ll 3 \times 10^{-5}$~s. 
It is worthy of mentioning that the angular frequency $\omega$ of long-wavelength Gregory-Laflamme type instability found for the Schwarzschild BH with the coupling $|\alpha| \gg r_h^2$ is in the range 
$\omega r_h \lesssim 0.1$ \cite{Held:2022abx}.
The time scale of this long-wavelength instability is larger than the order $10r_h$, so the Laplacian instability mentioned above destabilizes the Schwarzschild BH much more quickly. 
In particular, because of the peculiar dispersion relation $\omega \propto k^2$, the modes with larger values of $k$ are subject to Laplacian instabilities with shorter time scales.

We note that there are also mass terms for the dynamical perturbations $\vec{\psi}$. By looking at the equations of motion and setting to vanish the $r$-derivatives of $\vec{\psi}$, one finds that typical squared mass terms at large distances are of order ${\cal O}(1/|\alpha|)$ (as in the case of odd-parity perturbations). 
Therefore, the Laplacian instability occurs if the parameters satisfy the constraints $r_h^2\ll |\alpha|$ and $|\alpha|^{-1}\ll |\omega^2|\ll M^2$, where $M$ is a cutoff of the theory at most of order $\Mpl$.
Using the instability mode (\ref{ome4}), for instance, these conditions are expressed as
\be
|m_W^2|=
\frac{1}{2|\alpha|} \ll \frac{1}{r_h^2}\,,
\qquad {\rm and}\qquad
|m_W^2| \ll \frac{9\sqrt{13}}{52}
\frac1{r_h^2}\frac{(k r_h)^4}{l^3} \ll M^2\,.
\ee
Since we are considering the radial propagation with $k r_h \gg l \gg 1$, the latter inequality is satisfied for $r_h^{-1}\ll M$. 
For a BH with the solar mass $M_{\odot}=2 \times 10^{30}$~kg, this range corresponds to $r_h^{-1}=6.68\times10^{-20}~{\rm GeV} \ll M \lesssim \Mpl$, with $\sqrt{|m_W^2|} \ll 6.68\times10^{-20}$~GeV.

The finiteness of $m_W^2=1/(2\alpha)$ (for large values of $r$) is associated with the fact that the eigenvalues of ${\bm K}$ do not vanish for $r>0$. 
The same thing was also happening for odd-parity perturbations in current theory. However, this is in contrast with the behavior of the kinetic matrix of odd-parity perturbations in Einsteinian cubic gravity \cite{DeFelice:2023vmj}, where ${\rm det}({\bm K})$ approaches 0 at spatial infinity. 
The latter behavior is typically a signal of the strong coupling problem \cite{Pookkillath:2020iqq, BeltranJimenez:2020lee, Jimenez:2023esa}. 
In Einsteinian cubic gravity, this property also gives rise to a mass term of odd-parity perturbations growing as a function of $r$. 
In quadratic Weyl gravity, the strong coupling problem is absent for both odd- and even-parity perturbations. 
On the SSS background, we have shown that there are seven dynamical DOFs in total, whose number coincides with those obtained on the Minkowski and isotropic cosmological backgrounds \cite{Bogdanos:2009tn, Deruelle:2010kf, Salvio:2018crh}.
This fact also supports the absence of a strong coupling problem for the propagating modes.

It should be noticed that the non-Schwarzschild BH branch with the metric components (\ref{nS1}) is present for the quantity $p=r_h/\sqrt{2\alpha}$ in the range (\ref{prange}), i.e., 
\be
|\alpha| = {\cal O}(r_h^{2})\,.
\label{alrh}
\ee
This is the regime in which the WKB approximation starts to lose its validity. Hence our analysis based on the WKB approximation does not 
address the linear stability of such non-Schwarzschild BHs.
To see whether some instabilities arise for BHs in the coupling range $|\alpha| \lesssim r_h^{2}$, we need to resort to the numerical integration by setting proper boundary conditions for perturbations (in particular, $\bar{H}_1$) at spatial infinity and on the horizon.

\subsection{Angular speeds of propagation}

To study the BH stability along the angular direction, we consider solutions to the discriminant Eq.~(\ref{detAeq}) in the other limit $\omega\simeq l/r_{h}\gg k\gg r_{h}^{-1}$. 
In this case, there are four solutions with the dispersion relation $\omega \propto l$, where $\omega$ obeys
\be
\left( r^2 \omega^2-f l^2 \right) 
\left( \tilde\mu_1 \omega^6 + \tilde\mu_2 l^{2}\omega^4 
+ \tilde\mu_3 l^{4}\omega^2 +\tilde\mu_4 l^{6}
\right)=0\,,
\label{disc}
\ee
where $\tilde\mu_i$'s are $r$-dependent coefficients.
The angular propagation speed $c_{\Omega}=r{\rm d}\theta/{\rm d}\tau$, where $\tau$ is the proper time, can be expressed as $c_{\Omega}=r\omega/(\sqrt{f}l)$. 
For the Schwarzschild solution (\ref{Sch}) with $\beta=r/r_h$, the discriminant equation (\ref{disc}) reduces to 
\ba
& &
\left( c_{\Omega}^2 -1 \right) 
\{
81 \left[ 15+4\beta (2\beta-5) \right] c_\Omega^6
+9[795 + 4\beta(116\beta-343)] c_\Omega^4
-3[5697 + 4\beta (850\beta-2403)]c_\Omega^2 \nonumber \\
& &
+ 8865 + 20 \beta(272\beta-753)
\}=0\,.
\ea
In this equation, the coefficient of $c_\Omega^6$ never vanishes 
for real values of $\beta$. On the contrary, the last term, which does not have the $c_\Omega$ dependence, may vanish for finite values of $\beta$ in the range $\beta>1$.
However, that would only mean that we need to consider the next-to-leading-order coefficient ${\tilde\mu}_4$ in Eq.~\eqref{disc}.
As an example, let us consider the BH stability at the distance $\beta=r/r_h=2$. 
In this case, we have the following solutions
\ba
& &
c_{\Omega1,{\rm even}}^2=1\,,\label{cOe1}\\
& &
c_{\Omega2,{\rm even}}^2 \simeq -0.7291\,,\label{cOe2}\\
& &
c_{\Omega3,{\rm even}}^2 \simeq 1.1026-0.0761i\,,\label{cOe3}\\
& &
c_{\Omega4,{\rm even}}^2 \simeq 1.1026+0.0761i\,.\label{cOe4}
\ea
The first value (\ref{cOe1}) corresponds to the luminal propagation.
Since the second propagation speed squared (\ref{cOe2}) is negative, there is Laplacian instability for this mode.
The third and fourth modes have the following time dependence
\be
\psi_i \propto e^{-i\omega t} \propto \exp\!\left( \mp 
\frac{0.0128\,l}{r_h}t \right),
\ee
where the minus and plus signs correspond to Eqs.~(\ref{cOe3}) and (\ref{cOe4}), respectively, and we have neglected an oscillating term. 
While the amplitude of the third mode decreases in time, the fourth one exhibits exponential growth with oscillations. 
The time scale of Laplacian instabilities for the modes (\ref{cOe2}) and (\ref{cOe4}) can be estimated as $t_{\rm ins}={\cal O}(r_h/l)$ and 
$t_{\rm ins}={\cal O}(10^2 r_h/l)$, respectively.
For sufficiently large multipoles $l$, these instability time scales are less than $r_h$.

The angular Laplacian instabilities are physical for the parameters in the ranges $r_h^2\ll |\alpha|$ and $|\alpha|^{-1}\ll |\omega^2| \ll M^2$, where $M$ is the cutoff mass scale. 
On using the instability modes discussed above, these conditions translate to 
\be
|m_W^2|=
\frac{1}{2|\alpha|} \ll \frac{1}{r_h^2}\,,
\qquad {\rm and}\qquad
|m_W^2| 
\ll \frac{l^2}{r_h^2} \ll M^2\,,
\ee
where the latter gives the inequality $r_h^{-2} \ll M^2$. 
These conditions are the same as those discussed for radial propagating modes.

%%%%%%%%%%%%%%%%%%%%%%%
\section{Conclusions}
\label{consec}
%%%%%%%%%%%%%%%%%%%%%%%

In this paper, we have investigated the linear stability of BHs in quadratic Weyl gravity given by the action (\ref{Saction}). 
This theory generates the derivatives of metrics higher than 
second order in the field equations of motion. 
On the SSS background (\ref{background}) there are derivatives up to fourth order in metric components $f$ and $h$, but they can be eliminated to give the second-order differential 
Eqs.~(\ref{ddfeq}) and (\ref{ddheq}).
For any arbitrary Weyl coupling $\alpha$, the Schwarzschild background (\ref{Sch}) is always the solution to this theory. 
For the particular coupling range $0.876 \le r_h/\sqrt{2\alpha} \le 1.143$, there are also non-Schwarzschild hairy solutions where the metric components are approximately given by Eq.~(\ref{nS1}). 

In Sec.~\ref{oddsec}, we studied the propagation of metric perturbations in the odd-parity sector. 
Since the second-order action of odd-parity perturbations can be expressed as Eq.~(\ref{Sodd}) with the Lagrangian (\ref{Lag}), there are three dynamical perturbations $W$, $Q$, and $\chi$, where $\chi$ contains the second time derivative of $W$ as Eq.~(\ref{chidef}).
On using the WKB approximation with large values of the angular frequency $\omega$, wavenumber $k$, and multipole $l$, the equations of motion 
of dynamical perturbations reduce to the form (\ref{Aeq}) with (\ref{Aodd}).

We showed that there is at least one ghost mode in the odd-parity sector, but the strong coupling problem associated with a vanishing determinant of the kinetic matrix is absent in Weyl gravity. 
Moreover, independent of the radial distance $r$, the squared masses of dynamical DOFs are at most of the order $m_W^2=1/(2\alpha)$.
These properties are different from those in Einsteinian cubic gravity where the strong coupling problem leads to the blow-up of mass terms of odd-parity perturbations at large distances \cite{DeFelice:2023vmj}. 
We also found that, under the WKB approximation, the speeds of propagation of odd modes along the radial and angular directions 
are all equivalent to 1. 
Since we did not specify the background solution to derive these results, they are valid for both the Schwarizschild and non-Schwarizschild BHs. 
Moreover, we do not need to specify the range of the Weyl coupling 
constant $\alpha$ relative to $r_h^2$.

In the even-parity sector, we first introduced two Lagrange multipliers $\chi_1$ and $\chi_2$ to remove several higher-order time derivatives from the second-order perturbed action. 
Defining the rescaled fields $\bar{H}_1$ and $\bar{\chi}_2$ as in 
Eqs.~(\ref{barH1}) and (\ref{bchi2}), respectively, we showed that the second-order Lagrangian can be expressed as Eq.~(\ref{eq:canonical_L}) with four dynamical perturbations $\vec{\psi}=(h_1, \chi_1, H_2, \bar{\chi}_2)$. 
The Lagrangian also contains contributions of the nondynamical perturbation $\bar{H}_1$ and its radial derivative. On the background close to the Minkowski spacetime, the field $\bar{H}_1$ obeys the 
differential Eq.~(\ref{H1eq}), 
whose solution is given by Eq.~(\ref{H1sol}). 
To ensure the validity of the WKB approximation in the vicinity of the horizon, we require that the Weyl mass squared should be in the range $|m_W^2|r_h ^2 \ll 1$, i.e., $|\alpha| \gg r_h^2$. 
This is the regime in which only the Schwarzschild branch (\ref{Sch}) is present. 
Then, we investigated the linear stability of Schwarzschild BHs for the distance $r$ close to the horizon ($r_h^2 \lesssim r^2 \ll |\alpha|$). 

Among the four dynamical perturbations in the even-parity 
sector, there is at least one ghost mode. 
For large radial momentum modes with $k r_h \gg l \gg 1$, we found that one of the dispersion relations is given by $\omega=\sqrt{fh}\,k$ and hence its propagation is luminal. The other three dynamical perturbations satisfy the unusual dispersion relations $\omega \propto k^2$.
At the distance $\beta=r/r_h=2$ in the vicinity of the horizon, we showed that one of the squared propagation speeds $c_{r4,{\rm even}}^2$ is largely negative. This leads to the strong Laplacian instability whose time scale is much shorter than $r_h$. For the coupling $|\alpha| \gg r_h^2$, it is also known that there is a long-wavelength instability for Schwarzscild BHs \cite{Held:2022abx}. 
However, the time scale of this Gregory-Laflamme type instability is greater than the order $10r_h$, so the Laplacian instability found in this paper destabilizes the Schwarzscild BH much more quickly. 

For high angular momentum modes of even-parity perturbations ($l \gg k r_h \gg 1$), one of the dynamical DOFs propagates luminally. At the distance $\beta=r/r_h=2$, we found that two dynamical perturbations are prone to Laplacian instabilities, with time scales shorter than $r_h$ 
for large multipoles. 
We also note that the squared masses of even modes are at most of the order $m_W^2=1/(2\alpha)$ at large distances and that the strong coupling problem is absent in the even-parity sector.
Provided the model parameters are in the range $|\alpha|^{-1} \ll r_h^{-2} \ll M^2$, where $M$ is the cutoff mass scale at most of order $\Mpl$, the Schwarzschild BH is excluded by Laplacian instabilities both along the radial and angular directions. 

Several issues may deserve further investigation. First of all, we did not address the stabilities of Schwarzschild and non-Schwarzschild BHs for the coupling range $|\alpha| \lesssim r_h^2$ due to the limitation of the WKB approximation in the even-parity sector.
For this purpose, we need to solve the differential equation for $\bar{H}_1$ coupled with other dynamical perturbation equations of motion by giving proper boundary conditions on the horizon and at spatial infinity. It is also worth implementing the quadratic Ricci scalar term $\beta R^2$ in the action to see how the propagation of dynamical perturbations is modified. These issues are left for future work.

%%%%%%%%%%%%%%%%%%%%%%%%%%%%
\section*{Acknowledgements}
%%%%%%%%%%%%%%%%%%%%%%%%%%%%

The work of ADF was supported by the Japan Society for the Promotion of Science Grants-in-Aid for Scientific Research No.~20K03969 and by grant PID2020-118159GB-C41 funded by MCIN/AEI/10.13039/501100011033. 
ST is supported by the Grant-in-Aid for Scientific Research Fund of the JSPS No.~22K03642 and Waseda University Special Research Project No.~2023C-473.

\appendix

\section{Integrating the even-parity second-order action 
by parts}\label{app:by_parts}

To derive the second-order action of even-parity perturbations 
in the form (\ref{eq:canonical_L}), we perform the 
following integrations by parts:
\begin{enumerate}
\item $A{\dot\psi}''{\dot\psi}\to -A({\dot\psi}')^2+\frac12\,A''{\dot\psi}^2$\,,
    \item $A {\dot\psi}'{\dot\psi}\to-\tfrac12\,A'\,{\dot\psi}^2$\,,
    \item $A\psi''\psi\to-A\,(\psi')^2+\tfrac12\,A''\,\psi^2$\,,
    \item $A\psi''\psi'\to-\tfrac12\,A'\,(\psi')^2$\,,
    \item $A\psi'\psi\to-\tfrac12\,A'\,\psi^2$\,,
    \item $A{\dot\psi}\psi\to 0$\,,
    \item $A{\dot\psi}_1''{\dot\psi}_2 + B{\dot\psi}_2''{\dot\psi}_1\to
    -(A+B){\dot\psi}_1'{\dot\psi}_2'+\frac12 (A''+B''){\dot\psi}_1{\dot\psi}_2 -\frac12(A'-B')({\dot\psi}_1'{\dot\psi}_2-{\dot\psi}_1{\dot\psi}_2')$\,,
    \item $A{\dot\psi}_1\psi_2' + B{\dot\psi}_2\psi_1'\to
    \tfrac12\,(A+B)\,({\dot\psi}_1\psi_2'+{\dot\psi}_2\psi_1')+\tfrac14\,(A'-B')\,(\psi_1{\dot\psi}_2-\psi_2{\dot\psi}_1)$\,,
    \item $A\psi_1{\dot\psi}_2 + B\psi_2{\dot\psi}_1 \to
    \tfrac12\,(A-B)(\psi_1{\dot\psi}_2-\psi_2{\dot\psi}_1)$\,,
    \item $A{\dot\psi}_1'{\dot\psi_2} + B{\dot\psi_1}{\dot\psi}_2'\to
    \tfrac12\,(A-B)\,({\dot\psi}_1'{\dot\psi_2}-{\dot\psi}_2'{\dot\psi_1})-\tfrac12\,(A'+B'){\dot\psi_1}{\dot\psi_2}$\,,
    \item $A{\dot\psi}_1' \psi_2'' + B{\dot\psi}_2' \psi_1''\to
    \tfrac12\,(A+B)({\dot\psi}_1' \psi_2'' + {\dot\psi}_2' \psi_1'')
    -\tfrac14\,(A'-B')({\dot\psi}_1'\psi_2'-{\dot\psi}_2'\psi_1')$\,,
    \item $A \psi'''\psi''\to-\tfrac12\,A'\,(\psi'')^2$\,,
    \item $A \psi'''\psi'\to-A\,(\psi'')^2+\tfrac12\,A''\,(\psi')^2$\,,
    \item $A \psi'''\psi\to\tfrac32\,A'\,(\psi')^2-\tfrac12\,A'''\,(\psi)^2$\,,
    \item $A\psi_1'\psi_2+B\psi_1\psi_2'\to
    \tfrac12\,(A-B)(\psi_1'\psi_2-\psi_1\psi_2')-\tfrac12\,(A'+B')\psi_1\psi_2$\,,
    \item $A\psi_1''\psi_2'+B\psi_1'\psi_2''\to
    \tfrac12\,(A-B)(\psi_1''\psi_2'-\psi_1'\psi_2'')-\tfrac12\,(A'+B')\psi_1'\psi_2'$\,,
    \item $A\psi_1''\psi_2+B\psi_1\psi_2''\to -(A+B)\psi_1'\psi_2'
    -\tfrac12\,(A'-B')(\psi_1'\psi_2-\psi_1\psi_2')+\tfrac12\,(A''+B'')\psi_1\psi_2$\,,
    \item $A\psi_1'''\psi_2+B\psi_1\psi_2'''\to 
    -\tfrac12\,(A-B)(\psi_1''\psi_2'-\psi_1'\psi_2'')+\frac32\,(A'+B')\,\psi_1'\psi_2'
    +\tfrac12\,(A''-B'')(\psi_1'\psi_2-\psi_1\psi_2')-\tfrac12\,(A'''+B''')\psi_1\psi_2$\,,
    \item $A\psi_1'''\psi_2''+B\psi_2'''\psi_1''\to
    \tfrac12\,(A-B)(\psi_1'''\psi_2''-\psi_1''\psi_2''')-\tfrac12\,(A'+B')\,\psi_1''\psi_2''$\,,
    \item $A\psi_1''\psi_2'+B\psi_2''\psi_1'\to
    \tfrac12\,(A-B)(\psi_1''\psi_2'-\psi_1'\psi_2'')-\tfrac12\,(A'+B')\,\psi_1'\psi_2'$\,,
    \item $A\psi_1'''\psi_2'+B\psi_2'''\psi_1'\to
    -(A+B)\,\psi_1''\psi_2''-\tfrac12\,(A'-B')(\psi_1''\psi_2'-\psi_1'\psi_2'')+\tfrac12\,(A''+B'')\,\psi_1'\psi_2'$\,,
\end{enumerate}
where $A$ and $B$ are $r$-dependent functions.

\bibliographystyle{mybibstyle}
\bibliography{bib}

\end{document}